# Spatial self-phase modulation in the $H_2TPP(OH)_4$ doped in Boric Acid Glass

SrinivasaRao Allam[1], Mudasir H Dar[1], N Venkatramaiah[2], R Venkatesan[2], Alok Sharan[1*]

[1]Department of Physics, Pondicherry University, Puducherry - 605014, India

[2]Department of Chemistry, Pondicherry University, Puducherry-605014, India

*alok.phy@pondiuni.edu.in / aloksharan@gmail.com

Self-diffraction rings or spatial self-phase modulation (SSPM) was observed in tetra-phenyl porphyrin derivative 5,10,15,20 - *meso*-tetrakis (4-hydroxyphenyl) porphyrin ($H_2TPP(OH)_4$) doped in boric acid glass (BAG) at 671 nm excitation wave-length lying within the absorption band of sample with $TEM_{00}$ mode profile. Intensity modulated Z-scan was performed on these systems to study the thermal diffusion and to estimate the thermo-optic coefficients. The results obtained from self-diffraction rings experiment and modulated Z-scan are compared and analyzed for different concentration.

## 1. Introduction

Third order optical nonlinearity being the universally lowest order optical nonlinearity, are easily observable in all kinds of medium. Third order optical nonlinearity manifests in nonlinear absorption, optical Kerr effect, harmonic generation or nonlinear phase shift [1, 2]. Better understanding of these phenomena has led to wide variety of applications. Resonant systems exhibiting nonlinear absorption have their imaginary part of complex refractive index as intensity dependent and the onset of it is characterized by saturation intensity. $TEM_{00}$ Gaussian beams have spatially varying intensity across the beam profile. When these beams are incident on saturable absorbers it leads to the phase distortion of the beam wave-front thereby creating spatial self-phase modulation (SSPM). When the phase distortion is less than value π (pi), the SSPM causes self-focusing / defocusing. Self-focusing / defocusing is used to obtain the third order optical nonlinear coefficients by z-scan technique [3]. In the presence of phase distortion greater than π, the SSPM of single beam leads to self-diffraction rings around the propagation direction.

The formation of spatial ring based on their origin can be classified in two types viz: conical emission and self-diffraction rings. The re-emitted photon from the molecules/atoms follows conical direction so as to satisfy phase matching condition. This conical emission has been observed in dense metal vapors like barium, strontium, and sodium [4, 5]. In the presence of phase distortion greater than π, the spatial self-phase modulation of single beam leads to the self-diffraction rings around the direction of propagation. The diffraction rings originates due to thermo-optical refractive index gradient [6-13] formed in the material. We have observed position dependent, spatially modulated intensity fringes in the transmission geometry in sample of $H_2TPP(OH)_4$ doped in BAG with thickness of 68±5µm, 92±18µm and 67±5µm corresponding to molar concentrations of $5\times10^{-5}$M, $1\times10^{-4}$M and $4\times10^{-4}$M respectively. We have noticed these rings at incidence wavelength of 671 nm, which is about at 29 nm away from resonant peak at 700 nm having band-width of 66 nm. Self-diffraction process is thickness dependent process. The diffraction rings has been observed in medium with thickness typically in the range varying from mm to cm and only in very few cases the thermal rings has been studied in the thin films of thicknesses of few microns [14-23]. Our study involves sample of thickness in the range of 50-100 microns as mentioned above.

In section 2, we present the necessary theoretical formulation to understand the formation of refractive index gradient and show how it leads to the diffraction of laser beam in the nonlinear media creating self-diffraction rings. These have been further studied as a function of intensity and spot-size of the beam in the open aperture Z-scan. In section 3, we report our findings on thermo-optical coefficients obtained through intensity modulated z-scan. In the last section 4 we have compared the nonlinear refractive indices estimated from self-diffraction rings and intensity modulated Z-scan experiments.

## 2. Refractive index gradient

Materials possessing intensity dependent refractive index under cw laser illumination would lead to the formation of the refractive index gradient. The transverse refractive index variation follows transverse profile of the Gaussian beam as

$$n(r) = \begin{matrix} n_0 \pm n_2 I_0 \exp(-2r^2/\omega^2) & \text{if } \omega \leq r \\ n_0 & \text{if } \omega > r \end{matrix} \quad (1)$$

where, $I_0$ is the peak intensity of transverse beam profile, $\omega$ is the spot size and $n_2$ is the nonlinear refractive index of material. Non-linear refractive index can be expressed as $n_2 = \pm n_{2nr} \pm n_{2T} + i n_{2a}$ with $n_{2nr}$ being non-resonant refractive index, $n_{2T}$ as thermal nonlinearity and $n_{2a}$ as absorptive nonlinear refractive index.

The non-resonant refractive index ($n_{2nr}$) contribution is due to the dipole interaction of optical field with atoms/ molecules. The sign of the nonlinear optical coefficients determines the self-focusing / defocusing phenomena. Resonant contribution to the nonlinear refractive indices leads to absorptive nonlinear refractive index ($n_{2a}$) which may also contribute to thermal nonlinearity ($n_{2T}$). The absorptive nonlinear refractive index originates from the number of photons loss due to the absorption process. As seen in the figure 1 the excited species decays to ground state either through spontaneous emission ($A_{ij}$) or stimulated emission ($B_{ij}$). While the stimulated emitted photons follow the propagation direction of the incident beam, spontaneously emitted photon is in overall 4π degrees with equal probability. The fraction of photons due to spontaneous emission reaching the photo-diode is ΔΩ/4π, where ΔΩ is the solid angle subtended by cross-sectional area of detector. The remaining fractional loss of photons would be due to the absorption. The thermal contribution ($\kappa T = h\nu_1 - h\nu_2$) comes from the energy loss of each photon due to non-radiative decay of molecules from $S_1$ to $T_1$ transition as shown in the figure 1. The thermal refractive index is given as $n_{2T} = (dn/dT)\Delta T$ [24].

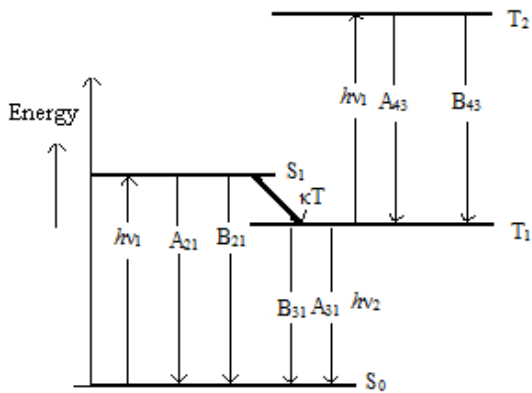

Figure 1. Four level diagram of H$_2$TPP(OH)$_4$: $S_1$ and $S_2$ are singlet states and $T_1$ and $T_2$ are triplet states. $h\nu_1$ is absorbing photon energy, $A_{ij}$ and $B_{ij}$ are spontaneous and stimulated coefficients.

The optical refractive index variation due to absorption follows TEM$_{00}$ Gaussian beam profile. This should also lead to similar thermal index profile due to rapid non-radiative decay. However this is perturbed or modulated due to thermal diffusion. The resultant refractive index gradient would diffract the incident beam leading to self-diffraction ring pattern. The refractive index gradient depth depends on the geometry of sample, diffusion coefficient, incident intensity and duration of irradiation of the sample. Our sample is sandwiched between two microscopic glass slides of 1.1±0.2 mm thickness with thermal conductivity 1.2 W/mK as shown in the figure 2. The thermal energy dissipated in the BAG from H$_2$TPP(OH)$_4$, diffuses through the glass slides to the surroundings at 21$^0$C room temperature.

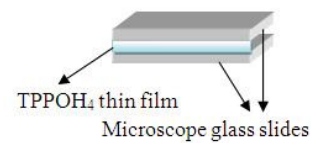

Figure 2. The geometrical structure of H$_2$TPP(OH)$_4$ doped in BAG sandwiched films of 50-100 μm thickness between two microscopic glass slides.

Spatial Self-Phase Modulation (SSPM) is the intensity modulation perpendicular to the direction of propagation of the emerging beam due to the transverse phase change because of the refractive index gradient. SSPM due to refractive index gradient was first time observed by W. R. Callen et al in the carbon disulfide with vanadium phthalocyanine dye [25]. The phase distortion of the beam profile in terms of nonlinear refractive index and radius of curvature of beam profile is given by

$$\Phi(r) = k_0 n(r) L + k_0 r^2 / 2R \qquad (2)$$

Where, $R$ is the radius of curvature of the beam profile and $k_0$ is free space propagation vector. Electrical field exiting the sample undergoes a

phase change during its propagation through the material, is given as

$$E(r, z_i + L) = E(0, z_i) \exp(-r^2/\omega_i^2) \exp(-\alpha_0 L/2) \exp(-i\Phi(r)) \quad (3)$$

The far field diffraction pattern of transmitted beam [12] profile is given by

$$I(\rho) = I_0 \left| \int_0^\infty J_0(k_0 \theta r) \exp(-r^2/\omega_i^2 - i\Phi(r)) r dr \right| \quad (4)$$

$$I_0 = 4\pi^2 \left| \frac{E(0, z_i) \exp(-\alpha L/2)}{i\lambda D} \right|^2 \quad \& \quad J(x) = \frac{1}{2\pi} \int_0^{2\pi} \exp(-ix \cos\varphi) d\varphi$$

Azimuthal symmetry of diffracted intensity distribution depends on the transverse profile of the incident beam. In our experiment we have seen circular symmetry in the diffraction pattern due to circular cross-section of Gaussian beam profile.

The refractive index variation in the radial direction is due to transverse nature of beam profile which further characterizes the diffraction pattern. As seen in figure 3, the ring thickness is decreasing in radially outward direction because the Gaussian beam profile intensity rapidly decreases in the transverse direction. We observed that there is particular threshold intensity for the ring formation and typically it takes one minute. The time taken to form the ring seems to decrease with increase of the incident intensity beyond the threshold intensity.

### 3. Z-Scan

#### 3.1 Open aperture z-scan

We performed the open aperture z-scan for the sample and simultaneously counted the number of diffraction rings in the transmitted intensity at different positions of the sample. Figure 3 shows the transmitted intensity profile of $H_2TPP(OH)_4$ doped in BAG for molar concentration of $4\times10^{-4}$ M at four different z-positions. Figure (3a) Corresponds to the position of the sample, where the incident intensity ($I=3.2\times10^4$ W/m$^2$) was below the threshold intensity of ring formation. We can see there is no diffraction pattern in the transmitted beam profile. Figure (3b) shows the diffraction pattern before the focus at position z = -2.8 mm for intensity $I=1.22\times10^6$ W/m$^2$. We can see the far-field intensity distribution of diffraction pattern is a series of thin diffraction rings with a central bright spot due to self-defocussing nature of the sample and convergent Gaussian beam [12, 20]. Figure (3c) corresponds to z=-0.2 mm with intensity $I=1.63\times10^8$ W/m$^2$. We notice that with decrease in the spot size, the central airy disc size has decreased. Figure (3d), is the ring pattern when the sample is positioned beyond the focus at z=0.3 mm with intensity of $I=1.21\times10^8$ W/m$^2$. The self-defocusing and divergence of beam produces far-field diffraction pattern containing the partially dark central Airy disk surrounded by thick diffraction rings. Images are recorded at the same screen distance for all four cases mentioned above in the far-field.

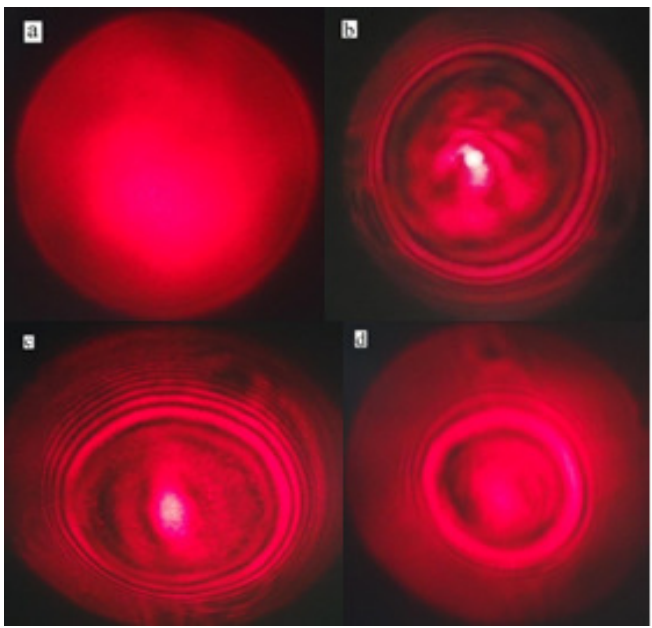

Figure 3. The transmitted intensity beam profile of $H_2TPP(OH)_4$ doped BAG at molar ratio $4\times10^{-4}$M: Irradiation (a) below the threshold intensity, (b) at I=$1.22\times10^6$ W/m$^2$, z=-2.8 mm, (c) at I=$1.63\times10^8$ W/m$^2$, z=-0.2 mm, (d) at I=$1.21\times10^8$ W/m$^2$, z=0.3 mm.

We calculated the beam spot-size for different positions along the z-axis and try to see the corresponding change in size of the central Airy disk diameter of the diffraction pattern. We found that the central Airy disk size of the diffraction pattern follows the same trend as laser beam spot size, as seen in the figure 4. The solid line corresponding to theoretical fit of $D(z)=D_0(1+z^2/z_0^2)^{1/2}$ where, $D_0$ is the diameter of the Airy disk at focus. Variations of central Airy disk size of diffraction rings for three different concentrations of $H_2TPP(OH)_4$ doped in the BAG follows the trend with that of beam spot size up to the focus but not after the focus [26]. The central Airy disk size has decreased with increasing concentration but resultant diffraction pattern size increases with concentration.

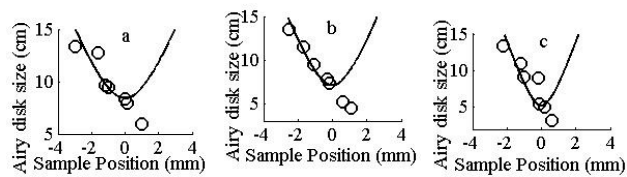

Figure 4.a, b and c are the Airy disk size variation with respect to position of the sample for the 5 x $10^{-5}$M, $1\times10^{-4}$M, and $4\times10^{-4}$M respective concentrations

The number of rings increases linearly [27] with incident intensity due to increase in the gradient of thermal refractive index. Figure 5 shows the variation of number of rings due to the sample position in the of Z-scan setup which essentially corresponds to intensity variation along the z-axis. The number of rings is fitted with $N(z)=N_0(1+z^2/z_0^2)^{-1}$ following the intensity variation along z-axis, where $N_0$ is the number of rings at focus. These rings were formed at incident intensities beyond the threshold intensities of $2.89\times10^5$ W/m$^2$, $2.25\times10^5$ W/m$^2$ and $1.35\times10^5$ W/m$^2$ for the dye concentrations $4\times10^{-4}$M, $1\times10^{-4}$M, and $5\times10^{-5}$M respectively. The threshold intensity increases with concentration because of proportional increase in yield of RSA (Reverse Saturable Absorption) process [28]. In our system, the molecule after absorption undergoes transition from $S_0$ to $S_1$ energy level. Some of them rapidly decay non-radiatively to $T_1$ state, which is meta-stable state and get trapped. This non-radiative decay produces thermal energy. $T_1$ to $T_2$ transition being energetically more favorable leads to reverse saturable absorption [29]. We notice increase in RSA behavior with increasing concentration. The increase in excited state absorption leading to the

RSA, decreases the number of photons for $S_0$ to $S_1$ transitions. As a result the number of non-radiative transitions from $S_1$ to $T_1$ decreases. Resultant thermal energy generation due to non-radiative decay decreases.

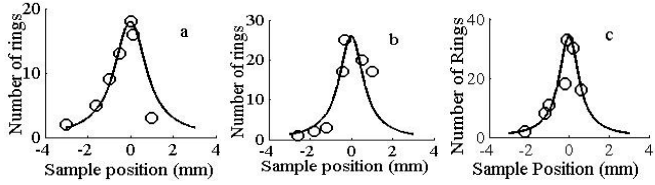

Figure 5. Variation of the number of rings with respect to z-position of the sample of $H_2TPP(OH)_4$ doped in BAG with (a) $5\times10^{-5}$ M, (b) $1\times10^{-4}$ M, and (c) $4\times10^{-4}$ M concentrations.

Table 1: Airy disc size and number of diffraction rings for sample positioned at beam waist.

| Concentration (M) | Airy disc size (mm) | Number of rings |
|---|---|---|
| $5\times10^{-5}$ | 8.4 | 18 |
| $1\times10^{-4}$ | 7.1 | 26 |
| $4\times10^{-4}$ | 5.2 | 34 |

We found that number of rings increased with concentration with simultaneous decrease in the size of Airy disk for same incident intensity. Table 1 contains the data of Airy disc size and number of diffraction rings at three different concentrations of $H_2TPP(OH)_4$ doped in BAG at beam waist position, where the spot size and intensity of illumination was kept constant for three concentrations. With increasing concentration, the Airy disc size was decreased and number of rings increased due increasing of thermo-optical nonlinear refractive index ($N = L\Delta n/\lambda$). The nonlinear refractive index as a function of number of diffraction rings is given by [27]

$$n_2 = \left(\lambda / 2n_0 IL\right) N \quad (5)$$

Here $n_0$ is linear real refractive index and for our sample was found to be at 1.45 with L being the sample thickness.

## 3.2 Closed aperture z-scan

In closed aperture z-scan the aperture size at the screen position was chosen to be equal to central Airy disc size to allow only Airy disc intensity. To understand the thermo-optical properties of $H_2TPP(OH)_4$ doped in BAG we have modulated the input intensity with equal duty cycle, by using optical chopper and varied the frequency of the chopper from few Hz to KHz frequencies. The transmitted pulses from the sample were collected by using Thor Labs Si-photodiode, having a nano second response time and were traced on the Agilent 5000A series oscilloscope. The screen shot is reproduced in the figure 6. ABCD is the original shape of the square pulse without the sample, and ABED is the transmitted pulse shape through the sample kept before the focus with incident intensity $2.6\times10^7$ W/m$^2$. The leading edge of the pulse has not changed because initially there is no thermal self-focusing. Under continuous illumination, the percentage of the optical energy contributing to the thermal energy increases with time and because of this thermal self-defocusing too increases. As a result we see the height of the pulse shape dropping from curve BC to BE. The transmitted pulse height decreased with time and after a particular time asymptotically reaches the point 'E' where the self-defocusing saturated.

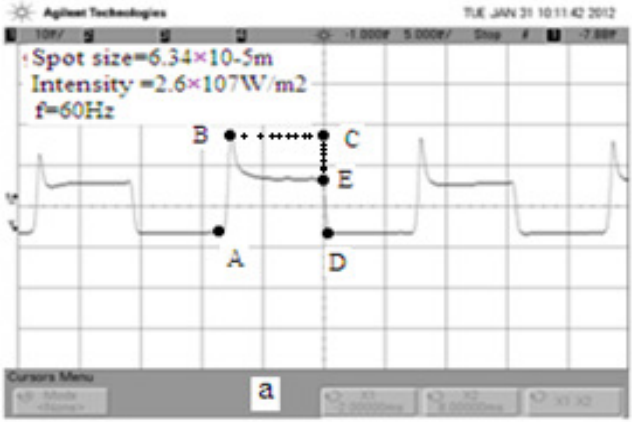

Figure 6. Snap shot of transmitted square pulse generated by chopping from the $H_2TPP(OH)_4$ doped BAG at $5\times10^{-5}$M for 60 Hz chopper frequency.

In the closed aperture z-scan, the transmitted intensity through the aperture depends on the effective focal length of nonlinear lens (NLL) formed due to nonlinear response of the sample. The focal length of NLL at a given position on the z-axis is expected to be constant for purely optical nonlinearity. However the presence of thermo-optical nonlinearity, the focal length changes with time due to thermal diffusion. The focal length of NLL formed due to thermo optical nonlinearity [30] is given by

$$f(t) = \frac{\pi \kappa \omega^3}{2.303 \rho A (dn/dT)} \left(1 + \frac{t_c}{2t}\right) \quad (6)$$

Where $\rho$ is the density of the material and the corresponding transmitted pulse intensity through the aperture as a function of time is given by [30]

$$I(t) = I(0)\left\{1 + \frac{2.303 EA}{1+\left(t_c/2t\right)} + \frac{1}{2}\left[\frac{2.303 EA}{1+\left(t_c/2t\right)}\right]^2\right\}^{-1} \quad (7)$$

As shown in the figure 7, the transmitted pulse decay has been fitted with equation 7. The thermal diffusion constant (D) and temperature dependent refractive index is estimated by the equations $D = \omega^2/4t_c$ and $(dn/dt) = E\lambda\kappa/P$, where E is the enhancement of the linear portion of the response, $t_c$ is transmitted pulse decay time, $\omega$ spot size, P incident power and $\kappa$ is Boltzmann constant. These values are tabulated in the table-2. The decay time increased with increasing concentration because of number of molecules contributing in the optical to thermal conversion process increased. Diffusion constants also increased with concentration.

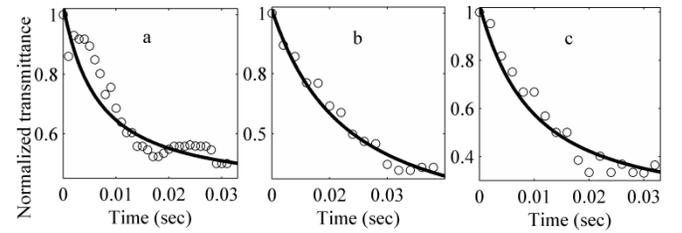

Figure 7. Transmitted pulse intensity of $H_2TPP(OH)_4$ doped BAG at (a) $5\times10^{-5}$M (b) $1\times10^{-4}$M (c) $4\times10^{-4}$ M for square pulse at 30 Hz with $2.6\times10^7$ W/m$^2$ average input intensity.

Table-2: Thermal properties of $H_2TPP(OH)_4$ doped BAG at three different concentrations

| $H_2TPP(OH)_4$ doped BAG (M) | Pulse decay constant (ms) | Diffusion constant ($10^{-5}$ cm$^2$/sec) | dn/dT ($10^{-19}$ /K) |
|---|---|---|---|
| $5\times10^{-5}$ | 60 | 1.2 | 3.2 |
| $1\times10^{-4}$ | 57 | 1.23 | 12.7 |
| $4\times10^{-4}$ | 33 | 2.16 | 24.3 |

Figure 8 shows the closed aperture z-scan of $H_2TPP(OH)_4$ at 50 Hz chopper frequency. The experimental data was fitted with equation 8 [31] valid for sample having large phase distortion to obtain the nonlinear refractive index. The nonlinear refractive indices obtained through closed aperture Z-scan (equation 8) and that used through number of self-diffraction rings (equation 5) are comparable and presented in table 3 below. In the processes of nonlinear refractive index measurement through number of rings, the error in the nonlinear coefficient is $\Delta n_2 = (\lambda/2n_0 IL)\Delta N$ i.e., the uncertainty in the $n_2$ is given by formation of ring by $\lambda$ (671 nm) path difference.

$$T(z) = \frac{1}{1 - \frac{4x\Delta\Phi_0}{(1+x^2)^2} + \frac{4x\Delta\Phi_0^2}{(1+x^2)^3}} \quad (8)$$

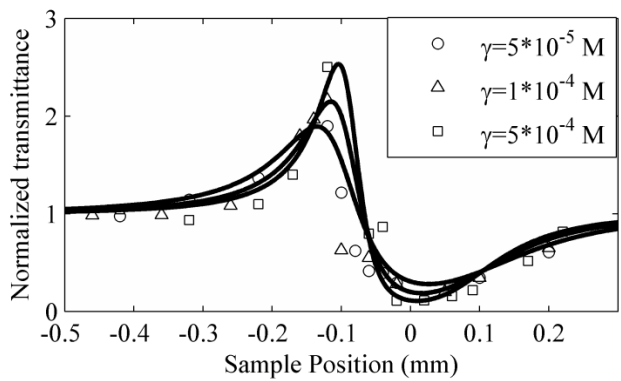

Figure 8. Closed aperture Z-Scan of $H_2TPP(OH)_4$ doped at BAG at 50 Hz chopper frequency for different concentrations.

Table-3: The nonlinear refractive index and susceptibility of $H_2TPP(OH)_4$ doped in BAG at three different concentrations

| Molar concentration (M) | Nonlinear refractive index (cm$^2$/W) from diffraction rings | Nonlinear refractive index (cm$^2$/W) by z-scan technique |
|---|---|---|
| $4 \times 10^{-4}$ | $1.4 \pm 0.4 \times 10^{-6}$ | $8.3 \times 10^{-7}$ |
| $1 \times 10^{-4}$ | $7.9 \pm 0.3 \times 10^{-7}$ | $6.4 \times 10^{-7}$ |
| $5 \times 10^{-5}$ | $7.3 \pm 0.4 \times 10^{-7}$ | $5.8 \times 10^{-7}$ |

## 4. Conclusion

We could systematically evaluate and understand self-diffraction/SSPM phenomenon in thin sample of typical 50-100 micron thickness, with phase distortion greater than $\pi$ in $H_2TPP(OH)_4$ doped BAG, caused due to passage of TEM$_{00}$ beam. The diffraction pattern obtained is well understood through the theoretical model proposed by L. Deng *et.al.,* [12]. The results obtained are in good agreement and compare favorably with that obtained from modulated Z-scan studies. The formation of refractive index gradient due to optical and thermal nonlinearities is also explained by four level model, which is good assumption for organic dye molecules. The number of diffraction rings linearly increased with intensity. The central Airy disk size of the diffraction pattern was seen to decrease with decreasing spot-size of the beam on the sample. With increasing concentration the size of central Airy disk decreases with simultaneously increase in the number of rings. The effect of RSA on the number of diffraction rings was explained successfully. From the time dependence of thermal lens effect, we are able to estimate the thermal decay constants and thermo-optic coefficients. The nonlinear refractive indices for different concentrations of $H_2TPP(OH)_4$ doped in BAG has been estimated from z-scan and nonlinear diffraction rings. The obtained results by these two methods are comparable. The lower value of the $n_2$ obtained through Z-scan is lower as the light through the thermal lens gets defocussed resulting in lesser light reaching the photodiode in closed aperture Zscan.